\documentclass[fleqn,preprintnumbers,a4paper,12pt,amsmath,amssymb,prb]{revtex4}
\usepackage{graphicx}
\usepackage{dcolumn}
\usepackage{bm}
\usepackage{epsf}
\newcommand{\newc}{\newcommand}
\newc{\beq}{\begin{equation}}
\newc{\eeq}{\end{equation}}
\newc{\beqa}{\begin{eqnarray*}}
\newc{\eeqar}{\end{eqnarray}}
\newc{\beqar}{\begin{eqnarray}}
\newc{\eeqa}{\end{eqnarray*}}
\newc{\bd}{\begin{displaymath}}
\newc{\ed}{\end{displaymath}}
\newc{\mbf}{\mathbf}
\begin{document}
\title{Quasi-solvability of Calogero-Sutherland model with Anti-periodic
  Boundary Condition}
\author{Arindam Chakraborty}
\affiliation{Dept. of Physics, Jadavpur University, Calcutta 700 032, India}
\author{Subhankar Ray}\email{sray@phys.jdvu.ac.in}
\affiliation{Dept. of Physics, Jadavpur University, Calcutta 700 032, India}
\affiliation{C N Yang Institute for Theoretical Physics, Stony Brook, NY
  11794, USA}
\author{J. Shamanna}
\email{jlsphy@caluniv.ac.in}
\affiliation{Physics Department, University of Calcutta, Calcutta 700 009, India}
\date{30 December 2006}
\begin{abstract}
The $U(1)$ Calogero-Sutherland Model with anti-periodic
boundary condition is studied. This model is obtained by applying a 
vertical magnetic field perpendicular to the plane of
one dimensional ring of particles. The trigonometric form
of the Hamiltonian is recast by using a suitable similarity
transformation. The transformed Hamiltonian is shown to be 
integrable by constructing a set of momentum operators which
commutes with the Hamiltonian and amongst themselves.
The function space of monomials of several variables
remains invariant under the action of these operators.
The above properties imply the quasi-solvability
of the Hamiltonian under consideration.  
\end{abstract}
\maketitle

\setcounter{equation}{0}
\setcounter{page}{1}

\section{Introduction}
The study of Calogero-Sutherland system 
has inspired significant research activity since the
pioneering work of Calogero and Sutherland \cite{cal62,suth71}. 
The integrability of the model has been studied 
for different root systems over the past few decades \cite{ols83}.
A few of the classical and spin varieties of the model are found to be 
exactly solvable and the solutions in terms of their eigenvalues and eigenfunctions
have been used extensively to describe physical properties of several
condensed matter systems. 
The study of Calogero systems is also related to various other research
areas in physics and mathematics, e.g.,
Yang-Mills theories \cite{gor94, mina94}, soliton theory \cite{poly95}, random
matrix model \cite{dyson62}, multivariable orthogonal polynomials 
\cite{jack69}, Selberg integral formula \cite{forr93}, $W^{\infty}$ algebra
\cite{hika93} etc.

This article investigates the Calogero-Sutherland Model (CSM) with anti-periodic
boundary condition. The anti-periodic boundary condition is a special case
of the general twisted boundary condition which arises when a one dimensional
chain of particles is placed in a transverse magnetic field.
A one dimensional chain of particles with a periodic boundary condition is
topologically equivalent to a one dimensional ring.
A particle transported adiabatically around this ring an integral number
of times returns to the same point.
In absence of a magnetic field this implies that the particle
returns to the same quantum state.
However, in the presence of a transverse magnetic field, one adiabatic transportation
around the ring introduces a phase factor $\exp(i \phi)$. 
This is called a twisted boundary condition.
When the phase factor is $\exp(i \phi) = -1$, it is called an 
anti-periodic boundary condition.
Though the introduction of a magnetic field is physically important in this
context, the model becomes mathematically more involved; and
the CSM with anti-periodic boundary condition remains less extensively
investigated.

The original version of the Calogero system incorporates 
long-range interaction by considering a two-body inverse square potential.
The integrability of such systems was initially studied by Calogero and 
Perelomov \cite{cal75, perel77} by means of Lax pair formulation. 
The integrability of CSM has since been 
investigated in a variety of ways \cite{ols83,mina93, berm97, poly99}.

The general form of CSM Hamiltonian is often represented
by the following equation:
\beq\label{hamilton}
H_N=\sum_{j=1}^N{\partial_j}^2-\lambda(\lambda-1)
 \sum_{\stackrel{\scriptstyle{j, k}}{j\neq k}} 
U(x_{jk}^-)
\eeq
The two-body potential, represented by $U(x_{jk}^-)$ is 
a long-range interaction in a chain of spinless nonrelativistic 
particles in one dimension. Here, $\lambda$ is a dimensionless interaction
parameter, $x_j$ and $x_k$ denote the coordinates
of the $j$-th and $k$-th particle respectively and $x_{jk}^-=x_j-x_k$.
While studying the solvability of $A_{N-1}$-type Calogero model, the Hamiltonian 
is operated on a partially ordered state space of all symmetric polynomials of 
several variables. 
This results in an upper triangular representation of the Hamiltonian. 
The diagonal terms of this matrix are the eigenvalues
of the Hamiltonian. 
The orthonormal eigenfunctions are expressed in terms 
of Jack symmetric polynomials \cite{jack69} which are very useful
in determining the various physical properties of many particle systems 
with long-range interactions \cite{habook}.

The search for an exact form of eigenfunction sometimes leads 
to partial diagonalization of the Hamiltonian \cite{tana05, fin01}.  
Among the one dimensional systems with periodic boundary condition, 
several such quasi-solvable models exist. The eigenvalues and eigenfunctions 
for many of them have been obtained\cite{tur87,ushbook}.
The model with $sl(2)$ structure was first discovered by Turbiner and
Ushveridze \cite{tur88}. 
It was also observed that the well known $N$ body Calogero-Sutherland models 
\cite{cal71,suth71,ruhl95} have similar Lie algebraic structure of $sl(N+1)$.

It may be noted that these models are in fact different generalizations
of the classically integrable Inozemtsev model \cite{tana04,ino83}. 
The common feature of these models with some 
underlying Lie algebraic structure is the existence of  
an invariant finite dimensional
module of the associated Lie algebra. Post and Turbiner \cite{post95}
studied a classification of linear differential operators of a single
variable which have a finite dimensional invariant subspace spanned by 
monomials.
One of the basic advantages of quasi-solvability
is that, one can restrict the study to a finite dimensional submanifold of
the full Hamiltonian. 
The finite dimensional matrix elements can be calculated by allowing the
Hamiltonian to act on finite-dimensional subspaces of a Hilbert space on 
which it is originally defined.
When the Hamiltonian operator preserves an infinite
number of subsequences of such finite dimensional subspaces \cite{tana05}
it becomes solvable.
The exact solvability of a model is ensured when the closure property is 
imposed on the space on which the Hamiltonian is allowed to act. 

In this article we study the integrability and solvability of
a spinless non-relativistic 
Calogero-Sutherland model (CSM) with anti-periodic boundary condition.
The two-body long-range interaction incorporating the anti-periodic boundary 
condition is derived. The Hamiltonian so obtained is reduced
to a more apparent integrable form, using a similarity transformation.
The integrability is then verified by constructing a set of mutually 
commuting momentum-like differential operators which further commute with 
the Hamiltonian.
Finally, the concept of quasi-solvability is discussed for a model
of many particle system. For CSM with anti-periodic boundary condition
the quasi-solvability
is studied by operating the Hamiltonian on a multivariable polynomial
space \cite{fin01}. The momentum operators in the anti-periodic model
remind us of the well known Dunkl operator \cite{dunk89, che91} 
which resembles the
Laplace-Beltrami-type operator acting on a symmetric Riemannian space.
These operators are extensively used in the study of integrability and 
solvability of Calogero-Sutherland models.
It is shown that these commuting momentum operators preserve the space spanned
by all monomials of degree $n$, i.e., $\{\prod_i z_i^{\ell_i}\}$, where 
$\ell_i\geq 0$ and $\sum \ell_i = n$, $n$ being a non-negative integer.
This property ensures the quasi-solvability of the Hamiltonian under study.

\section{Trigonometric version of CSM Hamiltonian}
Let us first consider the periodic CSM with inverse square long-range
interaction in the absence of a magnetic field. 
The topological representation of a one dimensional chain of particles with
periodic boundary condition is simply a circular ring.
A particle when transported adiabatically
around the ring an integral number of times, does not take up any phase
factor, and so the eigenfunctions retain their initial form.
Then the pairwise interaction summed around a unitarily equivalent circle
of circumference $L$, an infinite number of times is given as,
\beq\label{pair_int}
\sum_{n=-\infty}^{+ \infty} \frac{1}{(x+nL)^2}=\frac{1}{d(x)^2}
\eeq
where, as shown in the figure, $x$ is the interparticle distance along the 
ring and $d(x)$ is the chord length. It is easy to verify that 
$d(x) = L/\pi \sin(\pi x/L)$.
\begin{figure}[h]
\resizebox{!}{2.0in}
{\hskip 1cm \includegraphics{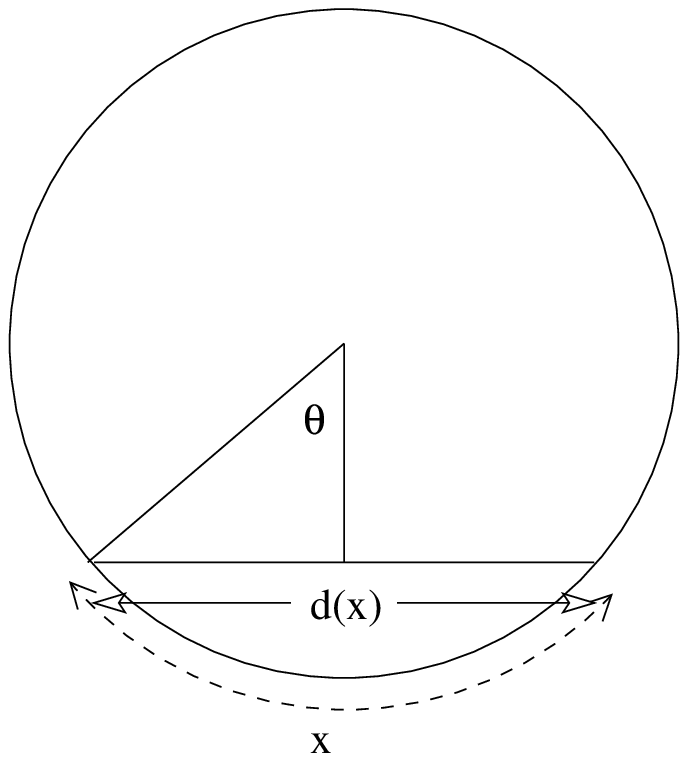}}
\resizebox{!}{2.0in}
{\hskip 1cm \includegraphics{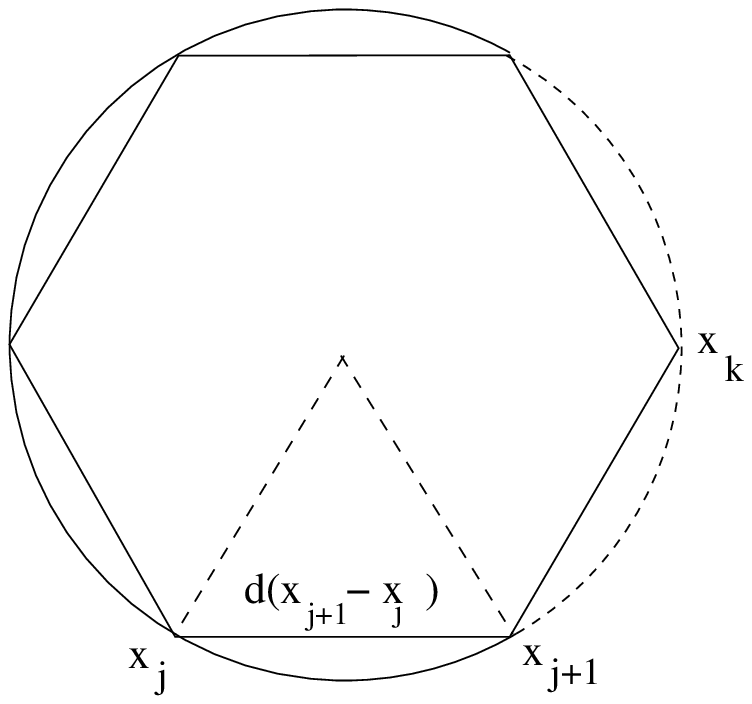}}
\caption{Inter-particle distances $d(x)$ and $d(x_{j+1}-x_j)$ for 
particles on a circular chain}
\end{figure}
Therefore, the potential $U(x)=(\pi^2/L^2)\sin^{-2}{x}$ is an inverse
trigonometric function of the inter-particle distance $x$.
The Hamiltonian with the above potential is given by,
\beq\label{htrig1}
H_N=\sum_{j=1}^{N}{\partial_j}^2-\lambda (\lambda-1)
\frac{\pi^2}{L^2} \sum_{\stackrel{\scriptstyle{j, k}}{j\neq k}} \frac{1}{\sin^2 (\pi x_{jk}^-/L)}
\eeq
where $x_{jk}^-= x_j-x_k$.
Using standard trigonometric identity, making a
change of variable $(\pi/2L)x_j \rightarrow x_j$ and
rescaling the Hamiltonian $(4L^2/\pi^2) H_N \rightarrow H_N $, 
Eq. (\ref{htrig1}) may be written as,
\beq\label{htrig2}
H_N=\sum _{j=1}^{N}{\partial_j}^2-\lambda
(\lambda-1)
\sum_{\stackrel{\scriptstyle{j, k}}{j\neq k}} \left(\frac{1}{\sin^2{(x_{jk}^-)}}+
\frac{1}{\cos^2{(x_{jk}^-)}}\right) .
\eeq
Let us now consider the anti-periodic case.
When a magnetic field is introduced transverse to the one
dimensional ring, a general twisted boundary condition arises.
A particle transported adiabatically around the entire
system $n$ number of times picks up a net phase $\exp{(in\phi)}$. 
The pairwise interaction summed around a unitarily equivalent circle
of circumference $L$, an infinite number of times, is now given as,
\beq\label{antipair_int}
\sum_{n=-\infty}^{+ \infty} \frac{\exp(i \phi n)}{(x+nL)^2} .
\eeq
The above summation can be performed by making the choice,
$\phi=2\pi p/q$, with $p, \,q$ relative primes, and 
$n=jq+k $, with $j, \, k$ integers ($-\infty < j < +\infty$ and 
$0 \le k \le q-1$).
The interaction term becomes,
\beq
\sum _{n=- \infty}^{+\infty}\frac{\exp{(i\phi n)}}{(x+nL)^2} =
\sum _{k=0}^{q-1} \sum _{j=- \infty}^{+\infty}
\frac{\exp(i 2 \pi pj) \exp(i 2 \pi p k/q)}{\left[(x+kL)+(qL)j \right] ^2}
=\sum _{k=0}^{q-1}\frac{\exp{(i 2\pi pk/q)}}
{\left[(qL/\pi) \sin{\frac{\pi(x+kL)}{qL}}\right]^2} .
\eeq
The last expression represents an interaction
with a general twisted boundary condition. The model can be viewed as
a system of interacting particles residing on a circle with circumference
$qL$. For $p/q =1/2$ the sum becomes,
\beq
\sum _{k=0}^{q-1}\frac{\exp{(i 2\pi pk/q)}}
{\left[(qL/\pi) \sin{\frac{\pi(x+kL)}{qL}}\right]^2} 
\Longrightarrow \sum _{k=0}^{1}\frac{\exp{(i\pi k)}}
{\left[(qL/\pi) \sin{\frac{\pi(x+kL)}{qL}}\right]^2}
=\sum _{k=0}^{1}\frac{(-1)^k}
{\left[(2L/\pi) \sin{\frac{\pi(x+kL)}{2L}}\right]^2} .
\eeq
This corresponds to an anti-periodic boundary condition \cite{habook}.
The potential then takes the following form
\beq\label{antiu}
U(x)= \frac{\cos{(\pi x/L)}}{(L^2/\pi^2)\sin^2{(\pi x/L)}}.
\eeq
Thus the Hamiltonian with anti-periodic boundary
condition, using standard trigonometric identity,
and scale changes $(\pi/2L)x_j \rightarrow x_j$ and
$(4L^2/\pi^2) H_N \rightarrow H_N $, becomes
\beq\label{htrig3}
H_N^{^{ap}}=\sum _{j=1}^{N}{\partial_j}^2-\lambda
(\lambda-1)
 \sum_{\stackrel{\scriptstyle{j, k}}{j\neq k}}
\left(\frac{1}{\sin^2{(x_{jk}^-)}}-
\frac{1}{\cos^2{(x_{jk}^-)}}\right) .
\eeq

\section{Integrability of the model Hamiltonian}
The integrability of such types of Hamiltonian is established
by constructing a complete set of commuting momentum operators
that also commute with the model Hamiltonian. These operators
were initially introduced in the study of Calogero-Sutherland model
with periodic boundary conditions (both spin and classical cases) 
and are known as Dunkl operators. Similar operators 
have been used to study Calogero-Sutherland-type models
derived from different root systems and are called Dunkl-type operators 
\cite{fin01, buch94}. 

To construct commutative Dunkl-type operators we introduce variables,
$z_j = \exp(2ix_j)$.
Using this substitution, the anti-periodic Hamiltonian becomes,
\beq\label{globe1}
H_N^{^{ap}}=\sum_{j=1}^N(z_j\partial_j)^2-\lambda(\lambda-1)
\sum_{\stackrel{\scriptstyle{j, k}}{j\neq k}}
\frac{z_j z_k}{(z_{jk}^-)^2}-\lambda(\lambda-1)
\sum_{\stackrel{\scriptstyle{j, k}}{j\neq k}}
\frac{z_j z_k}{(z_{jk}^+)^2} , \;\;\; \mathrm{where} \;\;\; z_{jk}^{\pm} = z_j \pm z_k .
\eeq
Let us apply the following similarity transformation,
\bd
\widetilde{H}_N=\Delta^{-1}H_N^{^{ap}}\Delta
\ed
where,
\bd
\Delta=\prod_{\stackrel{\scriptstyle{j, k}}{j < k}}\frac{(z_{jk}^-)^{\mu_1(\lambda)}(z_{jk}^+)^{\mu_2(\lambda)}}
{(z_j z_k)^
{(\{\mu_1(\lambda)+\mu_2(\lambda)\}/2)}} ,
\;\; \mathrm{and} \;\;
\mu_1(\lambda)=\lambda,1-\lambda \;, \;\;\;
\mu_2(\lambda) = \frac{1}{2}[1\pm\sqrt{1+4\lambda-4\lambda^2}] .
\ed
Thus, the anti-periodic Hamiltonian becomes
\beq\label{sim2}
\widetilde{H}_N=\sum_{j=1}^N(z_j\partial_j)^2
+ \frac{\mu_1(\lambda)}{2}
\sum_{\stackrel{\scriptstyle{j, k}}{j\neq k}}
\frac{z_{jk}^+}{z_{jk}^-}(z_j\partial_j-z_k\partial_k)
+ \frac{\mu_2(\lambda)}{2}
\sum_{\stackrel{\scriptstyle{j, k}}{j\neq k}}
\frac{z_{jk}^-}{z_{jk}^+}(z_j\partial_j-z_k\partial_k)
\eeq
The term $\mu_1(\lambda)$ is real for all $\lambda$, however, $\mu_2(\lambda)$ 
is real only for $1+4\lambda-4\lambda^2 \geq 0$, i.e., 
$\vert\lambda-\frac{1}{2}\vert \leq \frac{1}{\sqrt{2}}$. 
Under this restriction, the Hamiltonian $\widetilde{H}_N$ 
becomes hermitian. In the following, the integrability of the 
Hamiltonian is studied for different allowed values of $\lambda$. 

Let us introduce the coordinate exchange operators 
$\{\Lambda_{jk}\vert j,k = 1,..N; j \neq k\}$ and
the sign reversing operators $\{\Lambda_j\vert j,k = 1,..N\}$.
The coordinate exchange operator acting on the
coordinates of $j$-th and $k$-th particle may be defined by the
operation
$\Lambda_{jk}f(z_1,..,z_j,..,z_k,.., z_N)=
f(z_1,..,z_k,..,z_j,.., z_N)$.
This operator is (i) self-adjoint, (ii) unitary, and satisfies
(iii) $\Lambda_{ij}\Lambda_{jk}=\Lambda_{ik}\Lambda_{ij}
=\Lambda_{jk}\Lambda_{ik}$ ,
(iv) $\Lambda_{ij}\Lambda_{kl}=\Lambda_{kl}\Lambda_{ij}$ ,
(v) $\Lambda_{jk} z_k \partial_k = z_j \partial_j$ .

The sign reversing operator $\Lambda_j$ may be 
defined by its action on the coordinates of the $j$-th particle as
$\Lambda_j f(z_1,..,z_j,.., z_N)=
f(z_1,..,-z_j,.., z_N)$.
This operator is (i) self-inverse $\Lambda_{j}^{-1} = \Lambda_{j}$, 
(ii) mutually commuting $[\Lambda_j, \Lambda_k] = 0$, and satisfies
(iii) $[\Lambda_{ij},\Lambda_k]= 0$ ,  $i\neq j\neq k$,
(iv) $\Lambda_{ij}\Lambda_j=\Lambda_i\Lambda_{ij}$.

In terms of $\Lambda_{jk}$ and $\Lambda_j$, we introduce operator
$\widetilde{\Lambda}_{jk} = \Lambda_j \Lambda_k
\Lambda_{jk}$, which is (i) self-adjoint and (ii) unitary.
In addition it satisfies 
(iii) $\widetilde{\Lambda}_{ij}\widetilde{\Lambda}_{jk}
=\widetilde{\Lambda}_{ik}\widetilde{\Lambda}_{ij}
=\widetilde{\Lambda}_{jk}\widetilde{\Lambda}_{ik}$,
(iv) $\widetilde{\Lambda}_{ij}\widetilde{\Lambda}_{kl}
=\widetilde{\Lambda}_{kl}\widetilde{\Lambda}_{ij}$,
(v) $\widetilde{\Lambda}_{jk} z_k \partial_k = z_j \partial_j $.
In terms of the above mentioned operators, the Dunkl-type momentum 
operators $\{ D_j \vert j=1,\dots, N \}$ may be represented by the
following equation,
\beq\label{mom11}
D_j = z_j\partial_j
+\frac{\mu_1 (\lambda)}{2}\sum_{k(\neq j)}
\frac{z_{jk}^+}{z_{jk}^-}(1-\Lambda_{jk}) .
+\frac{\mu_2 (\lambda)}{2}\sum_{k(\neq j)}
\frac{z_{jk}^-}{z_{jk}^+}(1-\widetilde{\Lambda}_{jk})
\eeq
$ \widetilde{H}_N = \sum_j D_j^2 $.
These Dunkl-type operators commute with the coordinate exchange operators and the
operators $\widetilde{\Lambda}_{jk}$, i.e; $[D_j, \Lambda_{jk}] = 0$,
$[D_j, \widetilde{\Lambda}_{jk}] = 0$. They also commute among themselves and
because of the very nature of their construction, commute with the Hamiltonian;
$[D_j, D_k] = 0$, $[D_j, \widetilde{H}_N ] = 0$.
The existence of such an operator establishes the integrability of the system.

\section{Quasi-solvability of the model Hamiltonian}

The integrability does not necessarily imply
the existence of a function space involving the variables 
$\{z_j \vert j = 1,\dots, N \}$ such that $ \widetilde{H}_N $ can 
be represented in a diagonal form. However, sometimes it may so 
happen that operators like
$\widetilde{H}_N $, acting on a suitably chosen function subspace can preserve
the space partially. In such cases we introduce the term quasi-solvability.
A linear differential operator $H_N$ of several variables 
$\{z_j \vert j=1,\dots,N\}$, is said to be quasi-solvable if it preserves a
finite dimensional function space $V_{\nu}$ whose basis admits an analytic 
expression in a closed form i.e., 
\beq
H_N V_{\nu} \subseteq V_{\nu}, \hskip 1cm \textrm{dim} 
V_{\nu} = n(\nu) < \infty, \hskip 0.5cm \textrm{where} \hskip 0.5cm
V_{\nu} = \langle v_1(z),\dots, v_{n(\nu)}(z)\rangle. 
\eeq

One of the advantages of quasi-solvability is that one can explicitly
evaluate finite dimensional matrix elements $A_{kl}$ defined by 
\bd
H_N v_k = \sum_{l=1}^{n(\nu)}A_{kl}v_l, \hskip 1cm (k = 1,\dots, n(\nu)).
\ed
The finite dimensional submatrices $A_{kl}$ may be diagonalizable
even when the entire $H_N$ is not. 
If the space $V_{\nu}$ is the subspace of a Hilbert space on which
the operator $H_N$ is defined, the spectrum of $H_N$ can be computed 
algebraically, so as to obtain the exact eigenvectors of $H_N$ that belong
entirely to $V_{\nu}$. This is the typical nature of quasi-solvability.

A quasi-solvable operator is said to be solvable if the quasi-solvability
condition holds for an
infinite number of sequences of finite dimensional proper subspaces 
each containing its previous descendant. 
\bd
V_1 \subset V_2 \subset \dots \subset V_{\nu} \subset \dots
\ed  
Moreover, if the closure of $V_{\nu}$, as $\nu \rightarrow \infty $, is
the Hilbert space on which 
$H_N$ acts, we call $H_N$ to be exactly
solvable. 

Now, we shall show that the Dunkl-type momentum operators obtained 
in Eq.(\ref{mom11})
preserve the space $\mathcal{R}_n$ i.e., the space spanned by all monomials
of the form $\prod_i z_i^{\ell_i}$, where $\ell_i \geq 0$ and
$\sum_i \ell_i = n$, $n$ being a non negative integer. 

It is easy to verify that the operator $(z_j\partial_j)$ preserves the space
$\mathcal{R}_n$.
\beq
(z_j\partial_j) \prod_i z_i^{\ell_i} = \ell_j \prod_i z_i^{\ell_i}
\eeq
 
\noindent
We shall show that the second and third operators in Eq. (\ref{mom11})
preserve $\mathcal{R}_n$, i.e., 
$(z_j+z_k)/(z_j-z_k)(1-\Lambda_{jk})\prod_i z_i^{\ell_i}
\in \mathcal{R}_n $
and
$(z_j-z_k)/(z_j+z_k)(1-\widetilde{\Lambda})_{jk}\prod_i z_i^{\ell_i}
\in \mathcal{R}_n $.
They can be rewritten as 
\beq\label{ratio1}
\frac{z_j+z_k}{z_j-z_k}(1-\Lambda_{jk})\prod_i z_i^{\ell_i}
=\left(\prod_{i(\neq j,k)} z_i^{\ell_i}\right)(z_j+z_k)(z_jz_k)
^{\textrm{min}(\ell_j,\ell_k)}\textrm{sign}(\ell_j-\ell_k)
\sum_{r=0}^{\vert\ell_j-\ell_k\vert-1}z_j^{\vert\ell_j
-\ell_k\vert-1-r}z_k^r
\eeq
and
\beq\label{ratio2}
\frac{z_j-z_k}{z_j+z_k}(1-\widetilde{\Lambda})_{jk}\prod_i z_i^{\ell_i}
=\left(\prod_{i (\neq{j,k})} z_i^{\ell_i}\right)(z_j-z_k)(z_jz_k)
^{\textrm{min}(\ell_j,\ell_k)}\kappa(\ell_j,\ell_k)
\sum_{r=0}^{\vert\ell_j-\ell_k\vert-1}(-z_j)^{\vert\ell_j
-\ell_k\vert-1-r}z_k^r
\eeq
where 
$\textrm{sign}(0) = 0$ and $\textrm{sign}(\alpha) = \alpha / \vert \alpha \vert$
for $\alpha \neq 0 $.
And 
\bd
\kappa(\alpha,\beta) = 
\left\{ \begin{array}{cc}
1 & \hskip 1cm \alpha <\beta \\
0 & \hskip 1cm \alpha=\beta \\
-(-1)^{\alpha+\beta} & \hskip 1cm \alpha>\beta
\end{array} \right.
\ed
The right hand side of Eq.(\ref{ratio1}) can be expressed
as a sum of the following two terms, 
\beq\label{term1}
\left(\prod_{i\neq{j,k}} z_i^{\ell_i}\right)z_j (z_jz_k)
^{\textrm{min}(\ell_j,\ell_k)}\textrm{sign}(\ell_j-\ell_k)
\sum_{r=0}^{\vert\ell_j-\ell_k\vert-1}z_j^{\vert\ell_j
-\ell_k\vert-1-r}z_k^r
\eeq and 
\beq\label{term2}
\left(\prod_{i\neq{j,k}} z_i^{\ell_i}\right)z_k(z_jz_k)
^{\textrm{min}(\ell_j,\ell_k)}\textrm{sign}(\ell_j-\ell_k)
\sum_{r=0}^{\vert\ell_j-\ell_k\vert-1}z_j^{\vert\ell_j
-\ell_k\vert-1-r}z_k^r .
\eeq
Let $p_r^j$ and $p_r^k$ denote the powers of $z_j$ and $z_k$ in the $r$-th 
summand of Eq. (\ref{term1}).
Then $p_r^j = \max(\ell_j, \ell_k) - r $ and
$p_r^k = \min(\ell_j, \ell_k) + r$ .
Thus, $p_r^j + p_r^k = \ell_j + \ell_k$.
Therefore, the sum of powers of $z_j$ and $z_k$, in the $r$-th summand
is $(\sum_{i\neq j, k} \ell_i ) + \ell_j+ \ell_k = \sum_i \ell_i$.

Hence, the expression (\ref{term1}) is a member of $\mathcal{R}_n$. 
Similar calculation shows that the expression (\ref{term2})
also belongs to a space spanned by monomials of degree $n$. 
Thus, the second operator in Eq. (\ref{mom11}) preserves $\mathcal{R}_n$. 
In a similar manner it can be verified that the third operator in 
Eq. (\ref{mom11}) also preserves $\mathcal{R}_n$.
As the operators $\{D_j \vert j=1,\dots,N \}$ are linear and 
preserve the space $\mathcal{R}_n$, the Hamiltonian 
$\widetilde{H}_N(=\sum D_j^2)$ also preserves $\mathcal{R}_n$, and hence is 
quasi-solvable.

\section{Conclusion}
In this article we have studied the behaviour of one dimensional chain of 
particles in a magnetic field interacting through an inverse square potential. 
The anti-periodic boundary condition allows one to analyze a special form of
the above situation. The extension of the model to anti-periodic case reduces 
the algebraic symmetry of the root systems. This makes this model
mathematically more challenging. Here, we recast the Hamiltonian
to a new form by using a suitable similarity transformation. The transformed
Hamiltonian is shown to be integrable in the sense that there exists a 
complete set of commuting momentum operators which also commute with the
Hamiltonian. It is observed that the momentum operators are hermitian for
a certain range of the interaction parameter.

The new form of the Hamiltonian and its constituent momentum operators
indicate the existence of a multivariable polynomial space which is 
invariant under the action of the Hamiltonian. Indeed, it is observed 
that the momentum operators constructed in this article keep the 
monomial space $\mathcal{R}_n$ invariant. This invariance demonstrates
the quasi-solvability of the model. 

\section*{Acknowledgment}

AC wishes to acknowledge the Council of Scientific and
Industrial Research, India (CSIR) for fellowship support.
%
%
%

\end{document}